\documentclass{IEEE_lsens}
\usepackage{textcomp}
\usepackage{graphicx}
\usepackage{subcaption}
\usepackage{array}
\newcolumntype{C}[1]{>{\centering\arraybackslash}p{#1}}
\usepackage{epsfig}
\usepackage{epstopdf}
\usepackage{amsmath,graphicx}
\usepackage{tikz}
\usepackage{cite}
\usepackage{multicol}
\usepackage{cleveref}
\usepackage{float}

\ifCLASSINFOpdf
\else
\fi
\usepackage[T1]{fontenc} 
\usepackage{amsmath}
\usepackage[cmintegrals]{newtxmath}
\usepackage{bm}

\usepackage{array}
\usepackage{url}
\ifCLASSINFOpdf
\else
\fi
\providecommand{\hypersetup}[1]{\relax}

\hypersetup{pdftitle={Bare Demo of IEEE\_lsens.cls for IEEE Sensors Letters},
pdfsubject={Typesetting},
pdfauthor={Ayush Tripathi},
pdfkeywords={Class, IEEE, IEEE\_lsens, IEEE Sensors Letters, LaTeX, Typesetting, TeX}}

\begin{document}


\IEEELSENSarticlesubject{IEEE Sensors Letters}

\title{SCLAiR : Supervised Contrastive Learning for User and Device Independent Airwriting Recognition}

\author{\IEEEauthorblockN{Ayush Tripathi\IEEEauthorrefmark{1}, Arnab Kumar Mondal\IEEEauthorrefmark{2}, 
Lalan Kumar\IEEEauthorrefmark{1,3}, Prathosh A.P.\IEEEauthorrefmark{4}}
\IEEEauthorblockA{\IEEEauthorrefmark{1}Department of Electrical Engineering,  Indian Institute of Technology Delhi, New Delhi, India\\
\IEEEauthorrefmark{2}Amar Nath and Shashi Khosla School of Information Technology,  Indian Institute of Technology Delhi, New Delhi, India\\
\IEEEauthorrefmark{3}Bharti School of Telecommunication,  Indian Institute of Technology Delhi, New Delhi, India\\
\IEEEauthorrefmark{4}Department of Electrical Communication Engineering, Indian Institute of Science, Bengaluru, India\\
}}

\IEEEtitleabstractindextext{%
\begin{abstract}
Airwriting Recognition is the problem of identifying letters written in free space with finger movement. It is essentially a specialized case of gesture recognition, wherein the vocabulary of gestures corresponds to letters as in a particular language. With the wide adoption of smart wearables in the general population, airwriting recognition using motion sensors from a smart-band can be used as a medium of user input for applications in Human-Computer Interaction. There has been limited work in the recognition of in-air trajectories using motion sensors, and the performance of the techniques in the case when the device used to record signals is changed has not been explored hitherto. Motivated by these, a new paradigm for device and user-independent airwriting recognition based on supervised contrastive learning is proposed. A two stage classification strategy is employed, the first of which involves training an encoder network with supervised contrastive loss. In the subsequent stage, a classification head is trained with the encoder weights kept frozen. The efficacy of the proposed method is demonstrated through experiments on a publicly available dataset and also with a dataset recorded in our lab using a different device. Experiments have been performed in both supervised and unsupervised settings and compared against several state-of-the-art domain adaptation techniques. Data and the code for our implementation will be made available at https://github.com/ayushayt/SCLAiR.
\end{abstract}

}

\maketitle

\section{Introduction}

\subsection{Background}
Airwriting may be defined as the process of writing letters in free space using unrestricted finger movements \cite{7322243,7322267}. It can be used to provide a user with a fast and touch-less input option which can be employed for applications in Human-Computer Interaction \cite{4154947}. The recognition of writing from motion sensors has garnered attention over the past few years and numerous algorithms have been proposed for the task \cite{amma2014airwriting, LIU2009657, kim2014efficient, 8272739, patil2016handwriting, alam2020trajectory, 7457172}. The studies involving the use of motion sensors can be broadly divided into two categories, the first involving the use of dedicated devices such as a wearable glove \cite{10.1145/2540048}, Wii remote \cite{xu2016air,li2021cross,xu2021novel }, smartphone \cite{li2018deep} and vision-based \cite{roy2018cnn}. This approach however, involves the user to carry an extra physical device which may be cumbersome for the users. To mitigate this, the second category of approaches aim to recognize writing movements by using wearable devices such as a ring worn on the index finger\cite{jing2017wearable}. smart-bands\cite{yanay2020air}. 

\subsection{Related Work}

There have been several attempts made at recognizing gestures of the palm by using wrist-worn devices \cite{10.1145/3264929, wen2016serendipity}. However, most of the work around this area has been focused on the setting wherein a stable, flat surface is used during the writing process \cite{graves2008offline, 7178375, lin2018show}. The presence of a flat surface for writing leads to stabilization of the hand, thereby decreasing noise and also provides a degree of feedback to the user. Close to the current work, in \cite{7444820} the authors explored the problem of airwriting recognition using a wrist-worn device. The study involved a single participant and Dynamic Time Warping (DTW) was used as a distance measure for classification thereby making the study user dependent. In \cite{yanay2020air}, the authors have proposed a Convolutional Neural Network based user independent framework in addition to a DTW based user dependent method for recognizing airwritten English uppercase alphabets. While all these methods are shown effective in airwriting recognition, they are limited in terms of user and device dependent nature of the studies performed hitherto. Such a system has the drawback of not generalizing to large-scale population due inter-subject variability in writing the same character. Furthermore, difference in the acquisition sensor characteristics across different devices may prove to be detrimental for the performance of airwriting recognition systems. Motivated by this, the current study explores a new paradigm for both user and device independent recognition of airwritten alphabets using signals obtained from 3-axis accelerometer and gyroscope sensors.

\subsection{Objectives and Contributions}

In this work, we explore a supervised contrastive learning \cite{khosla2020supervised} based architecture for airwriting recognition. The main idea behind the approach is that the hidden representations of signals for the same alphabet are expected to be similar to each other, while the representation corresponding to signals of different alphabets will be fairly different. Given this intuition, a 2-stage classification approach is adopted wherein, in the first stage, an encoder (with a projection head) is trained with using the contrastive loss. Subsequently, the projection head is discarded and a classification head is trained while keeping the encoder weights fixed. We validate the efficacy of the proposed approach by performing leave-one-subject-out experiments on $2$ different datasets - a publicly available dataset \cite{yanay2020air} (source dataset) and a dataset collected in the lab (target dataset). We also analyze the performance of the approach in two different settings, (a) an unsupervised setting in which the model is trained solely using the source domain data and evaluated on the target domain samples, and (b) a supervised setting in which the encoder is trained using source domain samples and the classifier head is fine tuned using target domain samples. We demonstrate that the supervised contrastive learning based approach leads to improved recognition accuracy on the target dataset, compared to other domain adaptation techniques - DANN \cite{dann} , DRCN \cite{drcn} and DeepJDOT \cite{deepjdot} in both supervised and unsupervised settings.

\vspace{-0.2cm}

\section{Proposed Method}

\begin{figure}[t!]
  \centering
  \centerline{\includegraphics[width=0.75\linewidth, trim=0 30 0 0, clip]{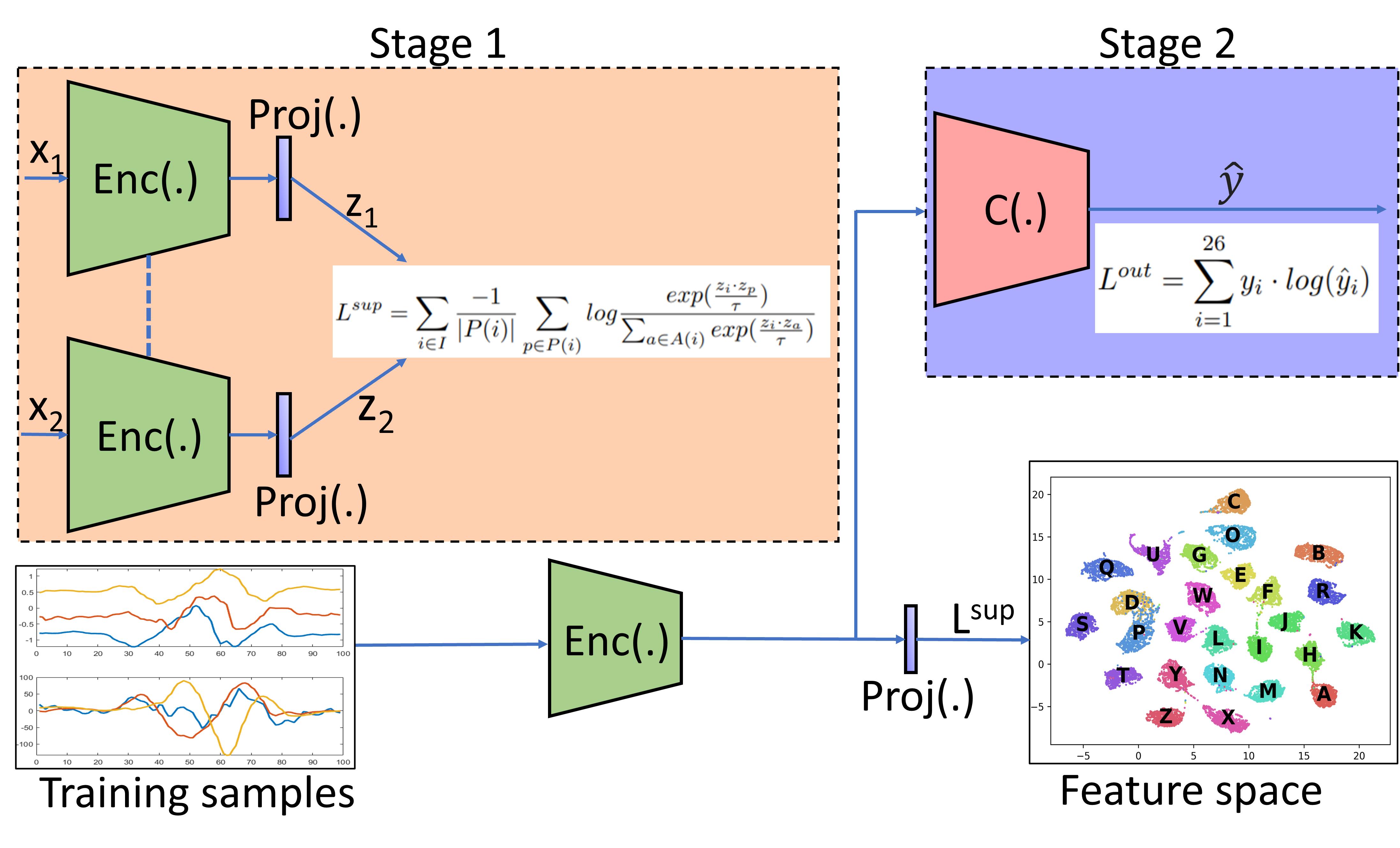}}
  \caption{Block diagram depicting our method. In the first stage an encoder with projection head based on deep network (called $Enc(\cdot)$ and $Proj(\cdot)$ respectively) is trained using supervised contrastive loss. Next, the projection head is discarded and a linear classifier ($C(\cdot)$) is cascaded on top of the feature learner and trained with Cross Entropy loss. The dashed lines indicate that the encoders have shared weights.} 
\label{fig:sup_con}
\vspace{-1.5em}
\end{figure}

The proposed framework consists of two stages, an encoder network trained using contrastive loss and a subsequent classifier head trained using the cross-entropy loss. Figure \ref{fig:sup_con} depicts the overall network. 
\vspace{-0.2cm}
\subsection{Problem Description}

A multivariate time-series recorded using 3-axis accelerometer and gyroscope of an Inertial Measurement Unit (IMU) placed on the wrist of the dominant hand of a human subject while writing uppercase English alphabets forms the input for the method. The signal is recorded over three axes (X, Y and Z), resulting in a total of six time series for each sample. Suppose we have a batch of $N$ sample and label pairs represented as $\{x_k,y_k\}_{k=1,2,..N}$  where $x_k \in \mathcal{X}$ is the input time series corresponding to the alphabet $y_k \in \{A,B,..Z\}$. The encoder network, $Enc(.) : \mathcal{X} \rightarrow R^{D_{E}}$ maps the input $x$ to a $D_E$ dimensional representation vector $r = Enc(x)$. Subsequently, the vector $r$ is normalized to unit hypersphere in $R^{D_E}$. There can be various choices of the encoder architecture that can be adopted without any constraints. Subsequently, the projection head maps the encoded representation $r$ to a vector $z = Proj(r) \in R^{D_{\mathcal{P}}}$. This comprises of the first stage of the training and involves learning the parameters of the encoder along with the projection head which is done by minimizing the supervised contrastive loss. It is to be noted that at the end of this stage of training, the projection network is discarded while the encoder weights are retained. 

The second stage is the classifier $C(.)$ which is a mapping from the latent space obtained from the encoder ($r$) to different alphabets. The predicted output is therefore given by $\hat{y} = C(r) = \sigma(W^Tr)$. Here, $W$ is the classifier weight matrix and $\sigma(.)$ is the softmax activation function. Since, the projection head is discarded after Stage $1$ of the training process, at inference-time, the overall model has the same number of parameters as a standard classification model trained using Cross-Entropy Loss. 

\subsection{Supervised Contrastive Loss}

Let $i\in I \equiv \{1,2,..N\}$ be index of an arbitrary sample from a batch of the training dataset, $P(i)$ be the set of indices of samples belonging to the same class as the $i^{th}$ sample in the batch and $A(i) \equiv I\setminus{\{i\}}$ be the set of indices of all samples from the batch other than the $i^{th}$ sample and $z_i = Proj(Enc(x_i))$ be the output of the projector head for input sample $x_i$. Supervised contrastive loss \cite{khosla2020supervised} is defined as,
\begin{equation}
    L^{sup} = \sum_{i\in I}^{} \frac{-1}{|P(i)|} \sum_{p\in P(i)}^{} log \frac{exp(\frac{z_i \cdot z_p}{\tau})}{\sum_{a\in A(i)}^{} exp(\frac{z_i \cdot z_a}{\tau})}
\end{equation}
In the above equation, $\tau$ is a scalar temperature parameter and $x \cdot y$ represents the inner product between vectors $x$ and $y$. Minimizing the aforementioned loss function encourages the encoder to learn closely aligned representation vectors for samples belonging to the same class by using the inner product as a measure of similarity between samples. The gradient of the loss function with respect to $z_i$ can be computed as 

\begin{equation}
   \frac{\delta L^{sup}_i}{\delta z_i}  = \frac{1}{\tau} \left\{ \sum_{p\in P(i)}^{} z_p \left(P_{ip} - \frac{1}{|P(i)|}\right)  + \sum_{n\in N(i)}^{} z_n P_{in} \right\}
\end{equation}
Here, $N(i)$ is the set of indices of samples not belonging to the same class as the $i^{th}$ sample in the batch, and $P_{ix} = exp(z_i \cdot z_x / \tau) / \sum_{a\in A(i)}^{} exp(z_i \cdot z_a /\tau)$. It can be shown that easier positive and negative pairs have small contribution towards the gradient compared to hard positive and negative pairs that have a larger contribution, giving rise to intrinsic hard positive/negative mining. 

\subsection{Model Architecture}

We explore the following different architectures for the Encoder block. Ablations on the hyperparameters for different encoder architectures can be seen in Figures \ref{fig:lstm_units} \& \ref{fig:num_filters} (Supplementary Material).

\begin{enumerate}
    \item 1DCNN : The architecture comprises of $4$ convolutional layers interleaved by a pooling layer, and followed by a Global Average Pooling layer. For the first couple of convolution layers, the number of filters is set to be $100$ while in the later stage at $160$, while the filter length for each of the layers is $10$. 
    \item LSTM and BiLSTM :  Here, vanilla LSTM and BiLSTM with $256$ number of units are used. 
    \item 1DCNN-LSTM/BiLSTM : Here, the input signal is passed through a couple of 1D Convolutional layers, followed by a pooling layer and the extracted features are then fed to a  $256$ unit LSTM/BiLSTM layer. The specifications of the convolutional layers are same as that of the first layers as used in the 1DCNN based encoder.
\end{enumerate}
The projection head is taken to be a single layer comprising of $128$ neurons, activated by the ReLU activation function. The parameters of the encoder and the projection head are learned by minimizing the supervised contrastive loss using the Adam optimizer. Further, the final classifier network is a single fully-connected layer comprising of $26$ neurons, and activated by softmax activation function with 50\% dropout. The classification network is trained using the usual cross Entropy Loss. The effect of varying the projection head dimension and the temperature parameter $\tau$ is presented in Figure \ref{fig:params} (cf. Supp.).

\section{Experiments and Results}

\subsection{Datasets}

\begin{itemize}
    \item \textbf{Source Dataset} - We use the publicly available dataset from \cite{yanay2020air} that consists of recordings obtained from $55$ subjects ($28$ female, $27$ male; $46$ right handed, $9$ left handed) while writing English uppercase letters ($15$ times). A dedicated app was built for the purpose of data collection which the users operated by using their non-dominant hand, while the signals were recorded from 3-axis accelerometer and gyroscope of a Microsoft band 2 worn on the dominant hand. The signals were recorded at the maximum sampling rate of $62$ Hz supported by the smart-band.
    \item \textbf{{Target Dataset}} - A dataset consisting of 3-axis accelerometer and gyroscope recordings of 20 subjects ($11$ male, $9$ female) was collected in a lab setting with consent of the participants. A Noraxon Ultium EMG sensor (having an internal IMU) \cite{noraxon} was placed on the wrist of the dominant hand of the participant who was then asked to write the English uppercase letters (repeated $10$ times). A user interface built using the Tkinter module in python was used for providing the participant with visual feedback for the letter to be written and also for automating the annotation process of the recordings. In Figure \ref{fig:data_collection} (Supplementary Material), we present the sample data collection setup. The signals were recorded at a sampling rate of $200$ Hz as supported by the IMU sensor and later downsampled to $62$ Hz in order to match to that of the source dataset. Our data will be made publicly available.
    
    Another target dataset comprising of recordings from $10$ subjects was recorded using a different device with sampling rate $400$ Hz (later downsampled to $62$ Hz) with the same experimental setup as described above.
    
\end{itemize}

\subsection{Experimental Details}
The recorded signals are obtained from different users writing at different speed and character sizes. Therefore the following preprocessing steps were employed.
\begin{enumerate}
    \item The samples are fixed to same length by padding zeros if the length of the sample is less than $L$ (taken to be 155 as in \cite{yanay2020air}), while discarding the extra samples otherwise.
    
    \item To the fix length samples, we apply the usual Z-score normalization for each of the $6$ individual signals. 
\end{enumerate}

The first set of experiments involves leave-one-subject-out (LOSO) validation on both the source and target datasets to avoid the user bias. The training data in each fold is split into a training set and a validation set having 80:20 ratio. A mini-batch training process with a batch size of $32$ is employed and early stopping with a patience of $5$ epochs.
\\ In the next set of experiments, Cross Entropy loss based classifier and Supervised Contrastive Loss based classifier have been compared in both supervised and unsupervised settings. In the unsupervised setting, the model is trained on the source dataset and evaluated on the target dataset. In the supervised case, the classifier head is fine tuned using the labelled target dataset in a LOSO fashion. The proposed approach has been compared with different domain adaptation techniques, by using the public implementations of the algorithms - DANN, DRCN and DeepJDOT. 

\begin{table}[!t]
\begin{minipage}[t]{0.49\columnwidth}
\caption{Mean recognition accuracy for leave-one-subject-out experiment on source dataset}
\label{tab:LOSO_source}
\centering
\scalebox{0.8}{
\begin{tabular}{lll}
\hline
\textbf{Architecture} & \textbf{CE} & \textbf{SCL} \\ \hline \hline
\textbf{1DCNN}        & 0.8441           & 0.8620                           \\  
\textbf{LSTM}         & 0.8298           & 0.8446                          \\ 
\textbf{BiLSTM}        & 0.8460            & 0.8760                           \\ 
\textbf{1DCNN-LSTM}   & 0.8480            & 0.8723                          \\
\textbf{1DCNN-BiLSTM}  & 0.8592           & 0.8805                          \\ \hline
\end{tabular}
}
\end{minipage}%
~~~
\begin{minipage}[t]{0.49\columnwidth}
\caption{Mean recognition accuracy for leave-one-subject-out experiment on target dataset}
\label{tab:LOSO_target}
\centering
\scalebox{0.8}{
\begin{tabular}{lll}
\hline
\textbf{Architecture} & \textbf{CE} & \textbf{SCL} \\ \hline\hline
\textbf{1DCNN}        & 0.8365           & 0.8619                          \\  
\textbf{LSTM}         & 0.7901           & 0.7928                          \\ 
\textbf{BiLSTM}        & 0.8015           & 0.8384                          \\ 
\textbf{1DCNN-LSTM}   & 0.8307           & 0.8455                          \\ 
\textbf{1DCNN-BiLSTM}  & 0.8269           & 0.8619                          \\ \hline
\end{tabular}
}
\end{minipage}
\vspace{-1em}
\end{table}

\

\begin{table}[!t]
\caption{Mean recognition accuracy on target dataset in the unsupervised setting}
\centering
\scalebox{0.75}{
\begin{tabular}{llllll}
\hline
\textbf{Architecture}  & \textbf{DANN} & \textbf{DeepJDOT} & \textbf{DRCN} & \textbf{CE} & \textbf{SCL}    \\ \hline\hline
\textbf{1DCNN}         & 0.7569                             & 0.7898                                 & 0.7583        & 0.7617      & \textbf{0.8030}  \\ 
\textbf{LSTM}          & 0.7519                             & 0.7463                                 & 0.7248        & 0.7440       & \textbf{0.7744} \\ 
\textbf{BiLSTM}        & 0.7619                             & 0.6557                                 & 0.7576        & 0.7840       & \textbf{0.8128} \\ 
\textbf{1DCNN-LSTM}    & 0.7336                             & 0.6809                                 & 0.7296        & 0.7296      & \textbf{0.7754} \\ 
\textbf{1DCNN\_BiLSTM} & 0.7575                             & 0.6588                                 & 0.7500        & 0.7757      & \textbf{0.8053} \\ \hline
\end{tabular}
}
\label{tab:unsupervised_results}
\vspace{-1em}
\end{table}

\begin{table}[!t]
\caption{Mean recognition accuracy on target dataset in the supervised setting}
\centering
\scalebox{0.75}{
\begin{tabular}{llllll}
\hline
\textbf{Architecture}  & \textbf{DANN} & \textbf{DeepJDOT} & \textbf{DRCN} & \textbf{CE}     & \textbf{SCL}    \\ \hline\hline
\textbf{1DCNN}         & 0.7775                             & 0.7921                                 & 0.7713        & 0.8240           & \textbf{0.8528} \\ 
\textbf{LSTM}          & 0.7746                             & 0.7273                                 & 0.7536        & \textbf{0.8178} & 0.8029          \\ 
\textbf{BiLSTM}        & 0.7909                             & 0.7151                                 & 0.7746        & 0.8207          & \textbf{0.8565} \\ 
\textbf{1DCNN-LSTM}    & 0.7685                             & 0.6853                                 & 0.7528        & \textbf{0.8344} & 0.8319          \\ 
\textbf{1DCNN\_BiLSTM} & 0.7928                             & 0.6463                                 & 0.7830        & 0.8590           & \textbf{0.8653} \\ \hline
\end{tabular}
}
\label{tab:supervised_results}
\vspace{-1em}
\end{table}

\begin{table}[!t]
\begin{minipage}[t]{0.49\columnwidth}
\caption{Mean accuracy on target dataset $2$ without fine-tuning}
\label{tab:nofinetuning}
\centering
\scalebox{0.8}{

\begin{tabular}{lll}
\hline
\textbf{Architecture} & \textbf{CE} & \textbf{SCL} \\ \hline \hline
\textbf{1DCNN}        & 0.9307         & 0.9581                           \\  
\textbf{LSTM}         & 0.8923           & 0.9007                          \\ 
\textbf{BiLSTM}        & 0.8850            & 0.9342                           \\ 
\textbf{1DCNN-LSTM}   & 0.9392            & 0.9476                         \\
\textbf{1DCNN-BiLSTM}  & 0.9076           & 0.9576                          \\ \hline
\end{tabular}
}
\end{minipage}%
~
\begin{minipage}[t]{0.49\columnwidth}
\caption{Mean accuracy on target dataset $2$ with fine-tuning}
\label{tab:finetuning}
\centering
\scalebox{0.8}{
\begin{tabular}{lll}
\hline
\textbf{Architecture} & \textbf{CE} & \textbf{SCL} \\ \hline\hline
\textbf{1DCNN}        & 0.9273           & 0.9573                          \\  
\textbf{LSTM}         & 0.9084           & 0.9130                          \\ 
\textbf{BiLSTM}        & 0.9088           & 0.9046                          \\ 
\textbf{1DCNN-LSTM}   & 0.9342           & 0.9388                          \\ 
\textbf{1DCNN-BiLSTM}  & 0.9331           & 0.9538                          \\ \hline
\end{tabular}
}
\end{minipage}
\vspace{-1em}
\end{table}

\subsection{Results and Comparison}

Table \ref{tab:LOSO_source} lists the performance of different model architectures by using both Cross Entropy (CE) and Supervised Contrastive Loss (SCL) for LOSO experiments with the accuracies averaged across all the subjects. It is seen that the SCL approach outperforms CE based approach for all model architectures. The results obtained by using this approach (Mean accuracy of 88.05\%) also outperforms the best reported accuracy $83.2\%$) on the given dataset \cite{yanay2020air}. It is also seen that the 1DCNN-BiLSTM model performs the best among all the model architectures. This may be attributed to the fact that the features extracted by the convolutional layer are beneficial for predicting the written alphabet. In Table \ref{tab:LOSO_target}, the results of LOSO experiments performed on the target dataset are tabulated with similar trends.

The evaluation of the models on the target domain samples in the unsupervised setting is presented in Table \ref{tab:unsupervised_results}. It is seen that in both CE and SCL based models, there is a decrement in the recognition accuracy when compared to the LOSO experiments in which both training and evaluation was done using samples from the target dataset. However, it can be seen that SCL based approach outperforms CE based approach in this scenario as well, thereby motivating the use of the approach for out-of-domain airwriting recognition system in case labeled target dataset is not available. The performance of the approach is also compared against prevalent domain adaptation techniques and it is observed that SCL approach yields superior performance.

\begin{figure}[t]
\begin{subfigure}{0.49\columnwidth}
  \centering
  \centerline{\includegraphics[width=\textwidth, trim=5 30 5 5, clip]{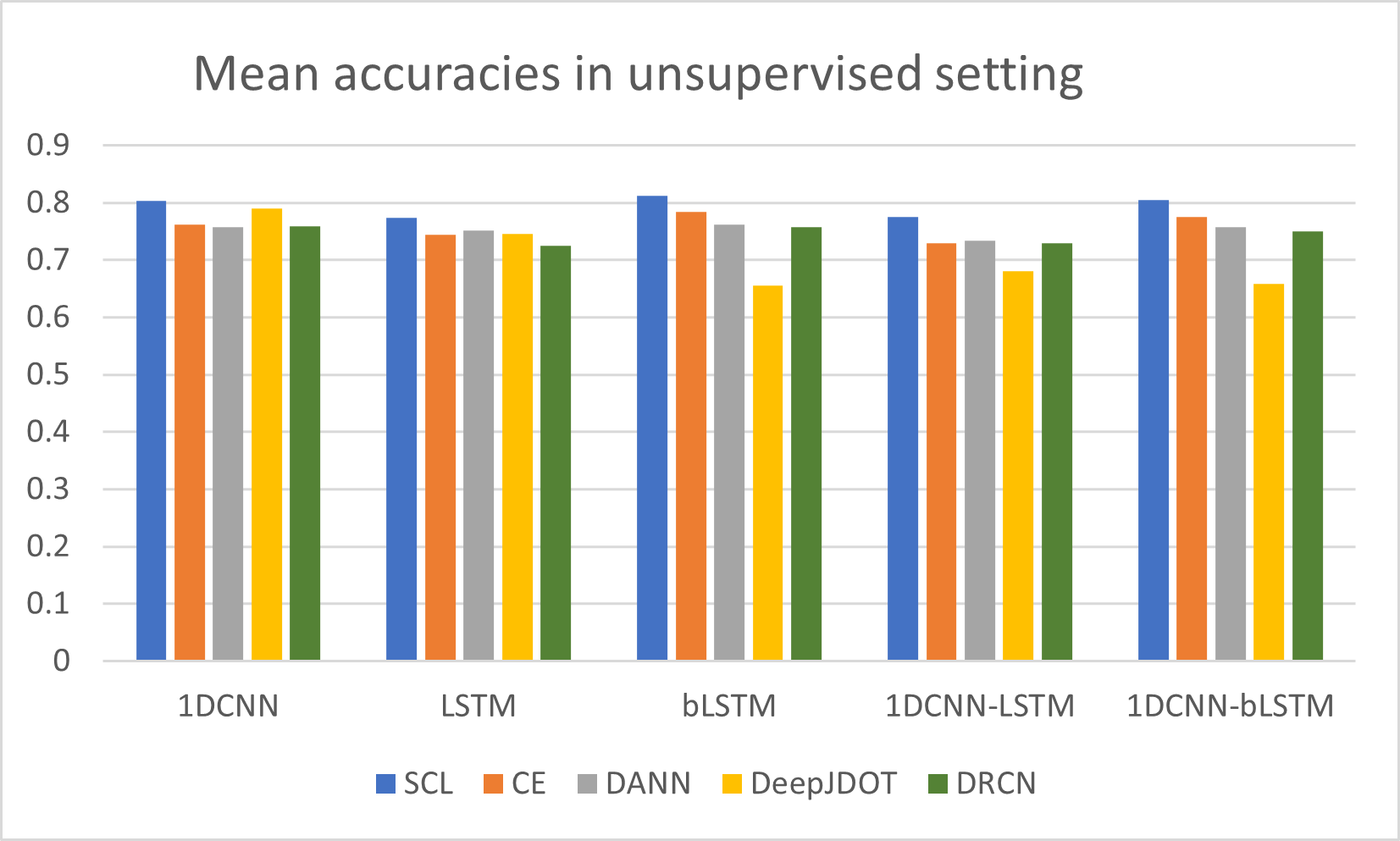}}
  \caption{}
  \label{fig:unsupervised}
\end{subfigure}%
~
\begin{subfigure}{0.49\columnwidth}
  \centering
  \centerline{\includegraphics[width=\textwidth, trim=5 30 5 5, clip]{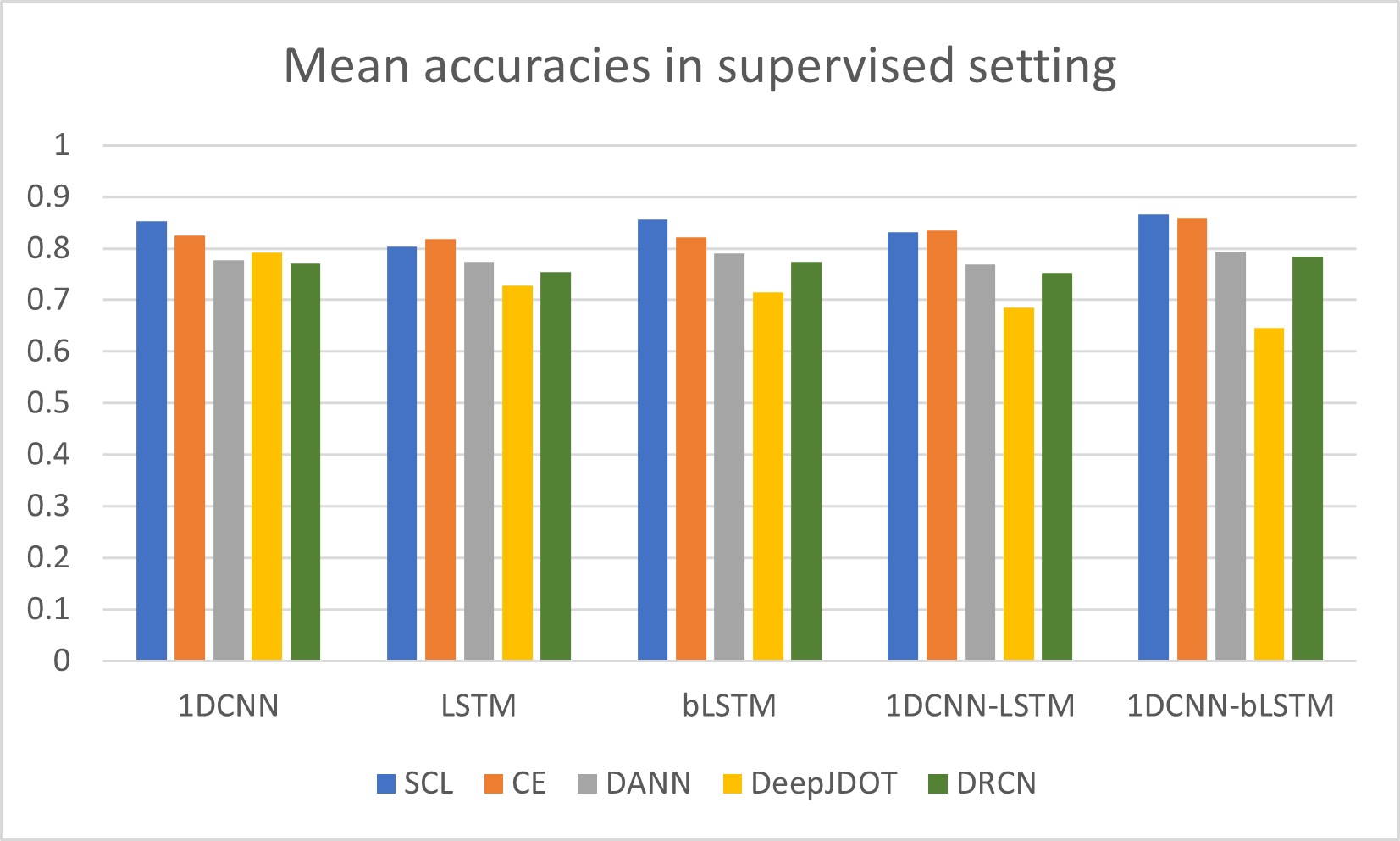}}
  \caption{}
  \label{fig:supervised}
\end{subfigure}
\caption{Mean accuracy on the target dataset using different transfer learning approaches in the (a) unsupervised and (b) supervised setting.}
\label{fig:sup_unsup}
\vspace{-1.5em}
\end{figure}

Furthermore, in Table \ref{tab:supervised_results}, we present the results for the set of experiments in which the classifier head is fine tuned using the labeled target dataset. As expected, the recognition accuracies are greatly improved when compared with the unsupervised setting. Also, it is observed that there is a marginal improvement from the LOSO case. The performance of CE and SCL based approaches are found to be quite close while outperforming the baseline domain adaptation techniques. The improvement in the recognition accuracies can be attributed to the fact that the fewer number of parameters are required to be learnt on tuning the classifier head in contrast to learning the parameters for the entire model. Therefore, the encoder parameters learnt on the relatively large source dataset along with the classifier parameters learnt on the small target dataset collectively boost the classification performance. While the trend of improved performance in supervised setting compared to the unsupervised case is in general valid for all the techniques, as illustrated in Figures \ref{fig:unsupervised} and \ref{fig:supervised} the performance of the fine-tuning approach is superior to DANN, DRCN and DeepJDOT. In \cref{tab:ConfusingLetters_LOSO_source,tab:ConfusingLetters_LOSO_target,tab:ConfusingLetters_unsupervised,tab:ConfusingLetters_supervised} (Supplementary Material), we present the top-5 most confusing letter pairs for the SCL based approach using the 1DCNN-BiLSTM architecture. In Tables \ref{tab:nofinetuning} and \ref{tab:finetuning}, we present the results of experiments performed by using different device on same subjects without and with fine-tuning. As it can be seen in the tables, SCL based approach outperforms CE based approach in a device-independent but user-dependent study.
\section{Conclusion}
In this paper, we explored supervised contrastive loss based framework for airwriting recognition using accelerometer and gyroscope signals obtained from a motion sensor worn on the wrist. It is seen that the proposed approach outperforms the state-of the-art accuracy on a publicly available dataset (source dataset), while also boosting the recognition accuracy on on unseen target dataset. It is seen that the supervised contrastive loss based approach outperforms existing domain adaptation techniques. This implies that our approach can be used for developing an airwriting recognition system, the performance of which is not greatly hampered by variations in the device and user. Future work will focus on exploring explicit methods of bias removal for dataset/user-style specific biases.

\bibliographystyle{IEEEtran}
\bibliography{refs.bib}

\clearpage

\section{Supplementary Material}

\subsection{Data Collection Setup}
\vspace{-0.2cm}

\begin{figure}[!ht]
    \includegraphics[width=.49\linewidth, keepaspectratio]{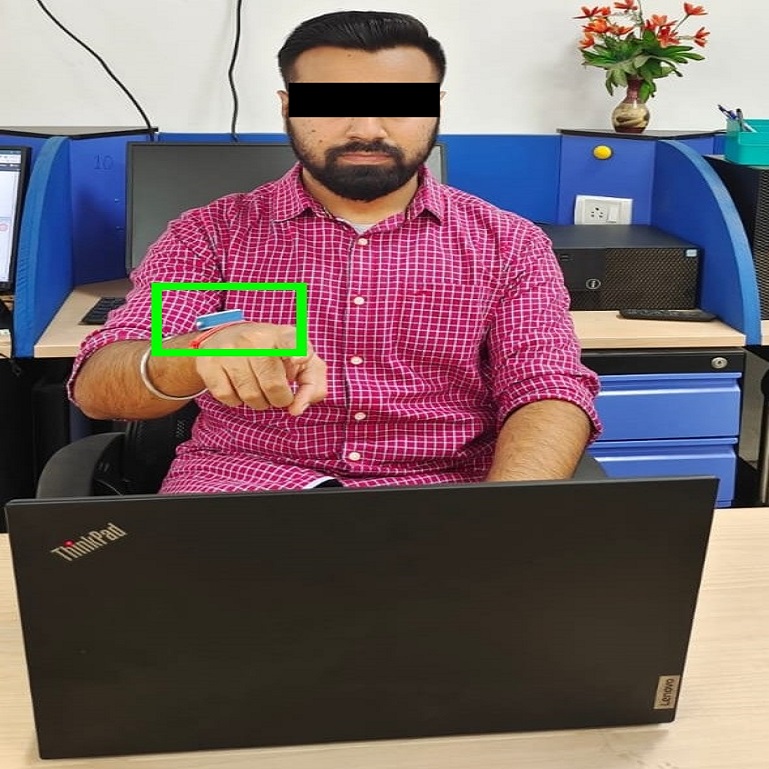}\hfill
    \includegraphics[width=.49\linewidth, keepaspectratio]{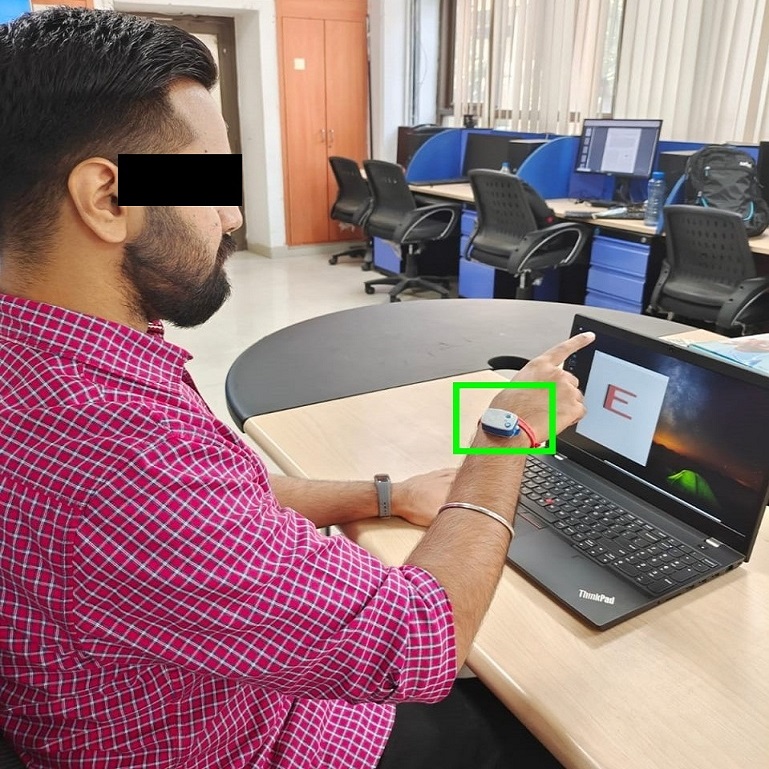}
    \caption{Depiction of the data collection setup. The IMU sensor placed on the wrist is depicted in the bounding box. The user interface operated by the subject's left hand is seen on the laptop screen.}
    \label{fig:data_collection}
\end{figure}

In this section, we present sample data collection setup depicted in Figure \ref{fig:data_collection}. The figure depicts front and side views of a male subject during the process of airwriting. The IMU sensor used to acquire accelerometer and gyroscope signals is placed on the wrist. For providing a visual feedback of the alphabet that is to be written and also automatically annotating the recordings, a user interface built using the Tkinter module in python was used.

\vspace{-0.6cm}

\subsection{Visualization of the Feature Space}
\vspace{-0.1cm}

In this section, we visualize the feature space vectors $z$, which are the output of the projection head. The $128$ dimensional vectors are reduced to $2$ dimensions by using Uniform Manifold Approximation and Projection (UMAP) \cite{mcinnes2018umap}.
\vspace{-0.5cm}
\begin{figure}[!h]
  \centering
  \centerline{\includegraphics[width=0.5\linewidth]{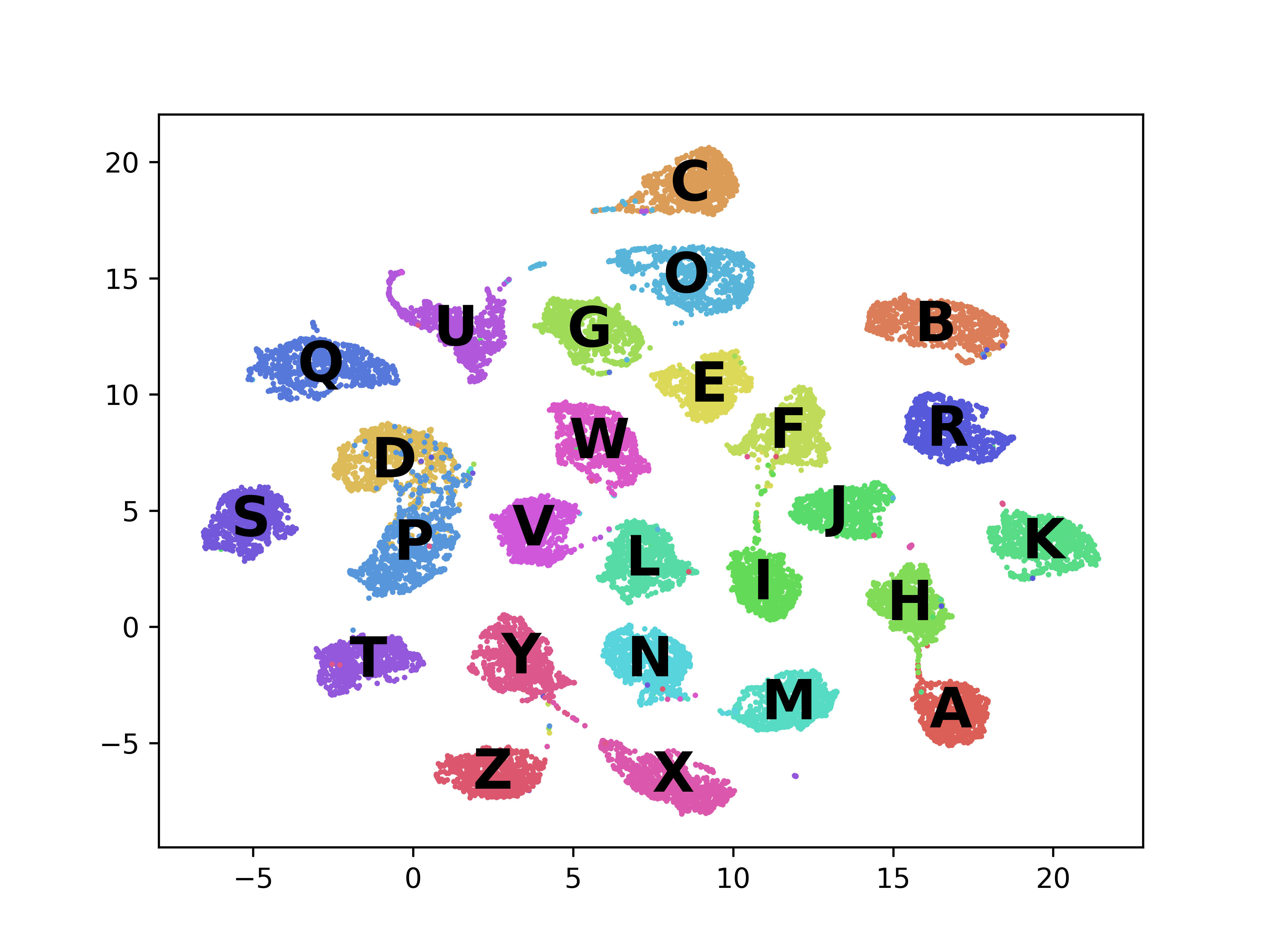}}
  \caption{Scatter plot depicting clusters corresponding to each of the $26$ alphabets obtained by applying dimensionality reduction using UMAP on the latent embedding vector}
\label{fig:clusters}
\end{figure}

\vspace{-0.8cm}

\subsection{Most Confusing Letter Pairs}
\vspace{-0.2cm}

In this section, we present the top-5 such pairs for different experiment settings in \cref{tab:ConfusingLetters_LOSO_source,tab:ConfusingLetters_LOSO_target,tab:ConfusingLetters_unsupervised,tab:ConfusingLetters_supervised}. The observations in these tables are also supported by Figure \ref{fig:clusters} as it can be seen that the clusters for the most confusing pairs are close to each other in the latent space.

\begin{table}[!b]
\begin{minipage}[b]{0.49\columnwidth}
\caption{The $5$ most confusing for LOSO experiments on source dataset.}
\label{tab:ConfusingLetters_LOSO_source}
\centering
\scalebox{0.75}{
\begin{tabular}{lll}
\hline
\textbf{Rank} & \textbf{Letters Pair} & \textbf{\% of Total Error} \\\hline\hline
1             & D,P                                       & 15.06\%                                        \\
2             & F,I                                       & 5.97\%                                         \\
3             & X,Y                                       & 3.75\%                                         \\
4             & N,W                                       & 3.39\%                                         \\
5             & C,O                                       & 2.34\%   \\ \hline                                     
\end{tabular}
}
\end{minipage}%
~~~
\begin{minipage}[b]{0.49\columnwidth}
\caption{The $5$ most confusing for LOSO experiments on target dataset.}
\label{tab:ConfusingLetters_LOSO_target}
\centering
\scalebox{0.75}{
\begin{tabular}{lll}
\hline
\textbf{Rank} & \textbf{Letters Pair} & \textbf{\% of Total Error} \\\hline\hline
1             & D,P                                       & 13.51\%                                        \\
2             & X,Y                                       & 4.87\%                                         \\
3             & G,O                                       & 4.45\%                                         \\
4             & N,W                                       & 4.32\%                                         \\
5             & G,Q                                       & 4.18\%   \\ \hline                                     
\end{tabular}
}
\end{minipage}
\end{table}

\begin{table}[!b]
\begin{minipage}[!b]{0.49\columnwidth}
\caption{The $5$ most confusing target pairs with the model trained on the source data}
\label{tab:ConfusingLetters_unsupervised}
\centering
\scalebox{0.75}{
\begin{tabular}{lll}
\hline
\textbf{Rank} & \textbf{Letters Pair} & \textbf{\% of Total Error} \\\hline\hline
1             & D,P                                       & 9.88\%                                        \\
2             & X,Y                                       & 7.61\%                                         \\
3             & G,Q                                       & 7.51\%                                         \\
4             & F,I                                       & 3.75\%                                         \\
5             & N,W                                       & 3.66\%   \\ \hline                                     
\end{tabular}
}
\end{minipage}%
~~~
\begin{minipage}[!b]{0.49\columnwidth}
\caption{The $5$ most confusing target pairs with a pre-trained model fine-tuned on target data}
\label{tab:ConfusingLetters_supervised}
\centering
\scalebox{0.75}{
\begin{tabular}{lll}
\hline
\textbf{Rank} & \textbf{Letters Pair} & \textbf{\% of Total Error} \\\hline\hline
1             & D,P                                       & 14\%                                        \\
2             & X,Y                                       & 5.86\%                                         \\
3             & N,W                                       & 4.43\%                                         \\
4             & G,Q                                      & 4.28\%                                         \\
5             & M,N                                       & 3.86\%   \\ \hline                                     
\end{tabular}
}
\end{minipage}
\end{table}

\subsection{Effect of Hyperparameters}
\vspace{-0.2cm}

\begin{figure}[!ht]
\begin{subfigure}{0.49\columnwidth}
  \centering
  \centerline{\includegraphics[width=\textwidth]{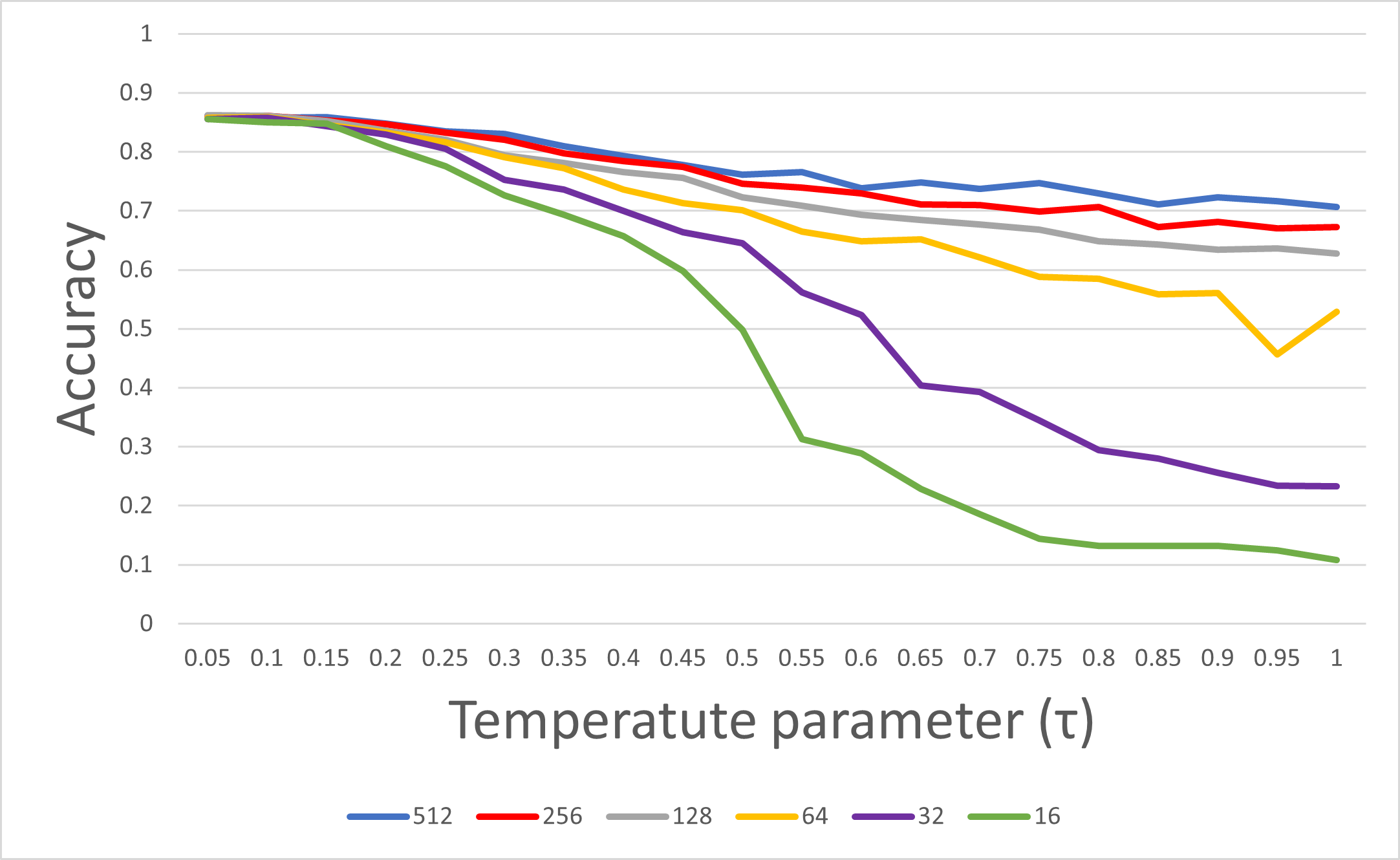}}
  \caption{Mean recognition accuracy vs. $\tau$ and $D_P$.}
\label{fig:params}
\end{subfigure}%
~
\begin{subfigure}{0.49\columnwidth}
  \centering
  \centerline{\includegraphics[width=\textwidth]{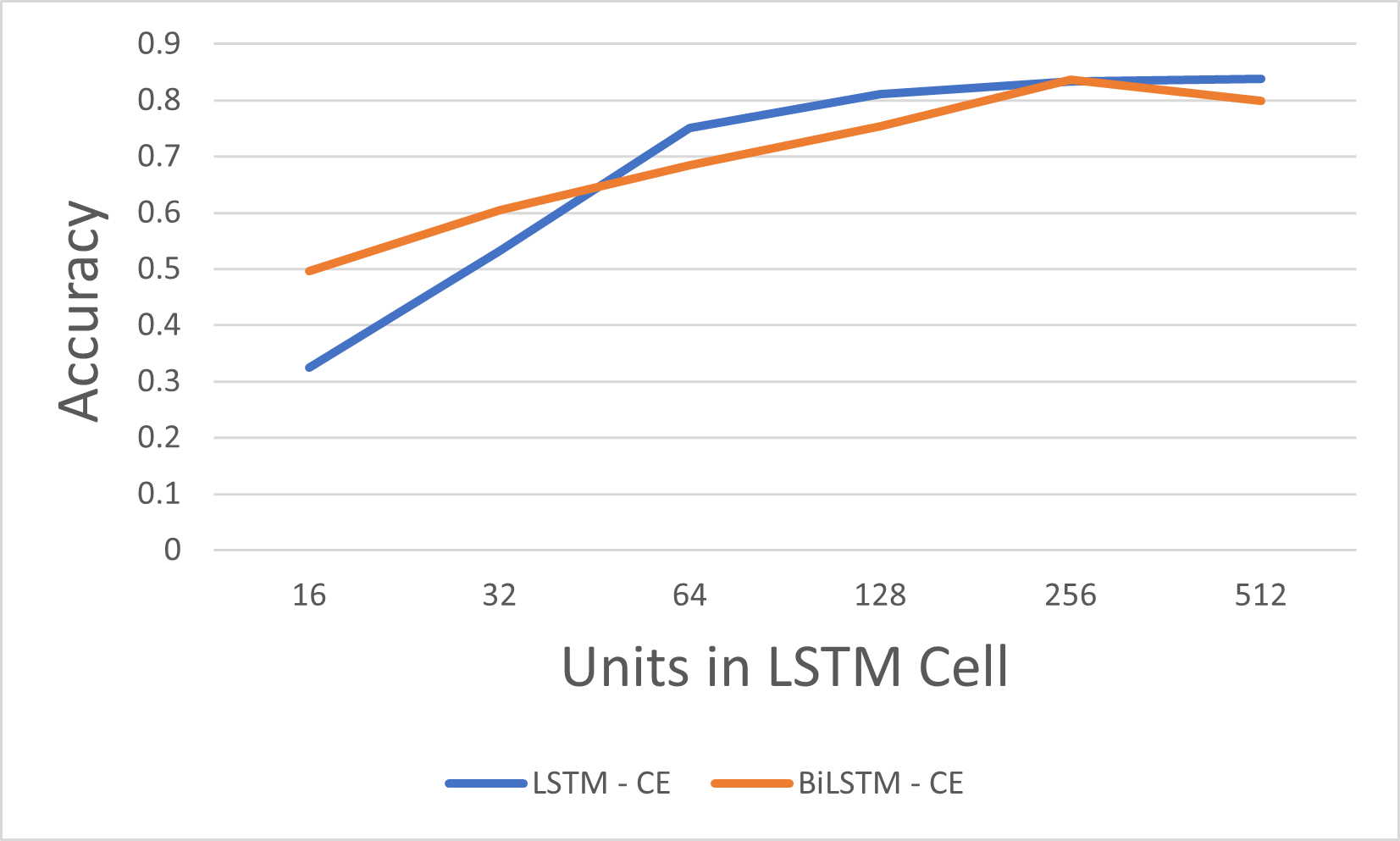}}
  \caption{Mean recognition accuracy vs. units in LSTM cells.}
\label{fig:lstm_units}
\end{subfigure}
\caption{Variation of mean recognition accuracy.}\label{fig:hyperparameters}
\end{figure}

\begin{figure}[!ht]
  \centering
  \centerline{\includegraphics[width=\linewidth]{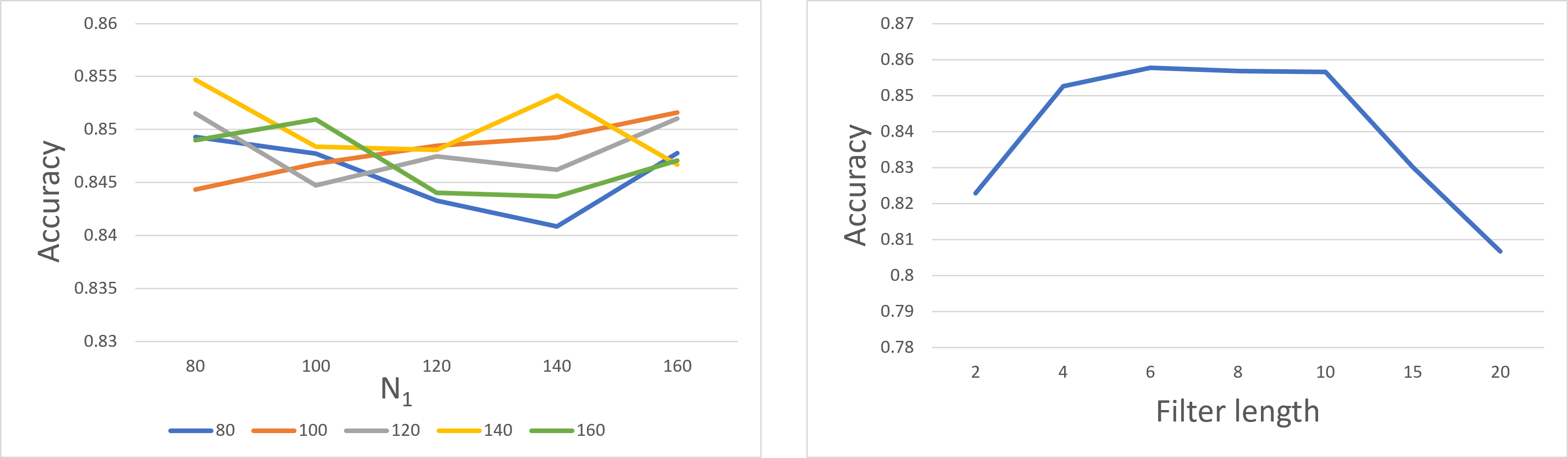}}
  \caption{Variation of accuracy by changing the number of filters (X-axis represents $N_1$ and the variation of $N_2$ is shown with different colors) and filter length for the 1DCNN architecture.}
\label{fig:num_filters}
\end{figure}

We analyze the effect of changing the parameters of the supervised contrastive loss on the classification performance. We changed the temperature parameter, $\tau \in \{0.05,0.1, .... ,0.95,1\}$ and the dimension of the projection encoding, $D_P \in \{512,256,128,64,32,16\}$. The mean accuracies achieved on LOSO experiments on the source dataset using the 1DCNN model are presented in Figure \ref{fig:params}. The variation of mean recognition accuracies with respect to the number of units in LSTM cell using CE loss for the LSTM and BiLSTM architectures are presented in Figure \ref{fig:lstm_units}. We also vary the number of filters for the first couple of convolution layers ($N_1$) and for the later layers ($N_2$), and the filter length, the effects of which on the classification performance are presented in Figure \ref{fig:num_filters}.

\subsection{Comparison with AdaBN}

In this section, we compare the proposed approach with AdaBN as in \cite{li2021cross}. For this task, we randomly selected 50\% of the subjects from the target dataset as a test set (2600 samples) and the remaining 50\% of the subjects from the target dataset were used for domain adaptation. We set up the experiment using the 1D-CNN architecture as described in the paper in both supervised and unsupervised settings and the results are presented in Table \ref{tab:adaBN}.

\begin{table}[H]
\caption{{Comparison of the proposed approach with AdaBN}}
\centering
\scalebox{1.0}{
\begin{tabular}{llll}
\hline
\textbf{Setting}  & \textbf{AdaBN} & \textbf{CE} & \textbf{SCL}     \\ \hline\hline
\textbf{Unsupervised}         & 0.6096                            & 0.7403     & 0.7723 \\ 
\textbf{Supervised}          & 0.7196                             & 0.7746     &  0.8046  \\ \hline                         
\end{tabular}
}
\label{tab:adaBN}
\end{table}

\end{document}